\pdfoutput=1
\documentclass[prl,twocolumn,superscriptaddress,showpacs]{revtex4}
\usepackage{graphicx}
\usepackage{amsmath}

\begin{document}

\title{Statistics at the tip of a branching random walk
and the delay of traveling waves }

\author{\'E. Brunet}
\author{B. Derrida}
\affiliation{Laboratoire de Physique Statistique, \'Ecole Normale Sup\'erieure, UPMC
Paris~6, Universit\'e Paris Diderot, CNRS, 24 rue Lhomond, 75005 Paris,
France}

\begin{abstract}
We study the limiting distribution of particles at the frontier of a
branching random walk. The positions of these particles can be viewed as
the lowest energies of  a directed polymer in a random medium in the
mean-field case. We show that the average distances between these leading
particles can be computed as the delay of a traveling wave evolving
according to  the Fisher-KPP front equation. These average distances
exhibit universal behaviors, different from those of the probability
cascades studied recently in the context of mean field spin-glasses.
\end{abstract}

\maketitle

The interest for branching random walks has a long history in Mathematics
\cite{McKean.75, Bramson.78, LalleySellke.87}, Physics and Biology. In
Biology they are commonly used to model the genealogies of evolving
populations, the spread of an advantageous gene or of an infection, the
combined effects of selection and mutations \cite{Fisher.37, Kessler.97,
GoldingKozlovsky.98}. In Physics  they  also appear in many contexts such
as reaction-diffusion models \cite{DoeringMuellerSmereka.03,
DeMasiFerrariLebowitz.86}, particle physics \cite{Munier.09,
Peschanski.08}, or the theory of disordered systems
\cite{DerridaSpohn.88, DeanMajumdar.01}.

In one dimension, the right frontier of a branching random walk is the
region located near its  rightmost particle.  An interesting question is
what does the branching random walk look like when seen from this
frontier. For example one can try to determine the position of the
second, the third, \dots or the $n^\text{th}$  rightmost particle in the
frame of the first rightmost particle. The statistical properties of
these positions depend on time and have a well defined long time limit
\cite{LalleySellke.87}  which we study in this letter using traveling
wave equations of the Fisher-KPP type \cite{Fisher.37, KPP.37, vanSaarloos.03}
\begin{equation}
{\partial h \over \partial t} = {\partial^2 h \over \partial x^2} + h - h^2.
\label{F-KPP}
\end{equation}

The fluctuating distances between  these rightmost particles allows one to
understand why directed polymers in a random medium \cite{DerridaSpohn.88}
have non-selfaveraging properties  similar to mean field spin-glasses
\cite{MPSTV.84}. Their study is also motivated by the growing interest for
the statistics of  extreme events \cite{TracyWidom.94, BouchaudMezard.97,
Burkhardt.07, Gyorgyi.08, MajumdarKrapivsky.02, SabhapanditMajumdar.07,
DeanMajumdar.01, Aizenman.05} which dominate a number of physical
processes \cite{IgloiMonthus.05, Kadanoff.06}. The last two decades have
seen the emergence of universal statistical properties of the probability
cascades describing the energies of the low lying states of several
spin-glass models \cite{Ruelle.87, BolthausenSznitman.98, Aizenman.05,
Aizenman.07, Arguin.07, BovierKurkova.06, BovierKurkova.07}. Somewhat
surprinsingly, as shown below, the distribution of the distances between
the extreme positions of particles in a branching random walk (which are
nothing but the energies of the low lying states in the mean field
version of directed polymer problem \cite{DerridaSpohn.88})  is different
from the predictions of the probability cascades \cite{Ruelle.87,
BolthausenSznitman.98, Aizenman.05, Aizenman.07, Arguin.07,
BovierKurkova.06, BovierKurkova.07}.

To start with a simple case, we consider a continuous time branching random
walk in one dimension: at time $t=0$  there is a single particle at the
origin $x=0$. This particle diffuses (for convenience we normalize the
variance of its displacement during time $t$ to be $2 t$) and branches at
rate~$1$. Whenever a branching event occurs, the offspring become
themselves branching random walks which diffuse and branch with the same
rates. The number of particles grows exponentially with time and
they occupy a region which grows linearly with time \cite{McKean.75,
Bramson.78}.

It has been known for a long time \cite{McKean.75, Bramson.78,
DeanMajumdar.01} that the distribution of the rightmost particle of
a branching random walk can be determined by solving a traveling wave
equation: the probability $Q_0(x,t)$  that, at time  $t$, there is no
particle at the right of of $x$,  satisfies
\begin{equation}
{\partial Q_0 \over \partial t} = {\partial^2 Q_0 \over \partial x^2}
+ Q_0^2 - Q_0 .
\label{Q0}
\end{equation}
[The derivation of (\ref{Q0}) is standard: one decomposes the
time interval $(0,t+dt)$ into two intervals $(0,dt)$ and $(dt,t+dt)$, and
write that
$Q_0(x,t+dt) =  Q_0(x,t)^2  dt +\langle Q_0(x-\eta,t) \rangle_\eta (1-dt) $
where the first term represents the contribution of a branching event and
$\eta$ in the second term the displacement due to  diffusion  during the first
time interval $(0,dt)$. With our
normalization $\langle \eta^2 \rangle= 2 dt$.] 

Up to the change $h=1-Q_0$, equation (\ref{Q0})   is the Fisher-KPP
equation (\ref{F-KPP}).
Since at time $t=0$ there is a single particle at the origin,  the initial
condition is simply 
\begin{equation}
Q_0(x,0)=1 \text{ \ for $x>0$}, \quad
Q_0(x,0)=0 \text{ \ for $x<0$}.
\label{initial-condition}
\end{equation}

If $Q_n(x,t)$ is the probability that there are exactly $n$ particles on the
right of $x$, one can see, as for $Q_0$, 
that the generating function $\psi_\lambda(x,t)$, defined as
\begin{equation}
\psi_\lambda(x,t) = \sum_{n \ge 0} \lambda^n Q_n(x,t),
\label{psi-def}
\end{equation}
evolves according to the same equation (\ref{Q0}), the only
difference being that the initial condition 
is replaced by
\begin{equation}
\psi_\lambda(x,0)=1 \text{ \ for $x>0$}, \quad
\psi_\lambda(x,0)=\lambda \text{ \ for $x<0$}.
\label{initial-condition-bis}
\end{equation}
We are now going to see that the knowledge of $\psi_\lambda(x,t)$  allows
one to obtain the average distances between the rightmost particles of
the system. If $p_n(x,t)$ is the probability of finding the $n^\text{th}$
rightmost particle at position $x$, it is easy to see that
\begin{equation}
 {\partial Q_0 \over \partial x}= p_1(x,t) \quad  \text{and } \ {\partial Q_n \over \partial x}=
p_{n+1}(x,t)  - p_n(x,t).
\label{pn}
\end{equation}
The average position $\langle X_n(t)\rangle$  
of the $n^\text{th}$ rightmost particle
 and the average distance $\langle d_{n,n+1}(t) \rangle$ between the
$n^\text{th}$ and $(n+1)^\text{th}$ rightmost particles can then be defined by
\begin{eqnarray}
\label{xn}
\langle X_n(t) \rangle = \int x\,   p_n(x,t) \  dx,     \\
\langle d_{n,n+1}(t) \rangle =  \langle X_n(t) \rangle - \langle X_{n+1}(t)
\rangle.
\label{dn}
\end{eqnarray}
[One should notice that the normalization of $p_n(x,t)$ is not $1$ but
$\int p_n(x,t) dx =(1-e^{-t})^{n-1}$ due to the events for which the total
number of particles at time $t$ is still less than $n$. One could prefer to
use different definitions of the positions or of the distances, for example
by conditioning on the fact that there are at least $n+1$ particles in the
system, but any such definition would coincide with (\ref{xn},\ref{dn}) up
to contributions which decay exponentially with time and disappear  in the
long time limit that we study below.]

With the definition (\ref{dn}) we obtain from
(\ref{psi-def},\ref{pn},\ref{xn}) that 
\begin{equation}
\sum_{n \geq 1}  \lambda^n 
 \langle d_{n,n+1}(t) \rangle = \int x \left[ {d Q_0 \over dx} - 
{d \psi_\lambda \over dx}  \right] dx,
\label{retard-1}
\end{equation}
which  relates the distances  $\langle
d_{n,n+1}(t) \rangle $  between the rightmost
particles  to the solution $\psi_\lambda(t)$ of the partial differential
equation (\ref{Q0}) with the
initial condition (\ref{initial-condition-bis}).

\begin{figure}[ht]
\includegraphics[width=\columnwidth]{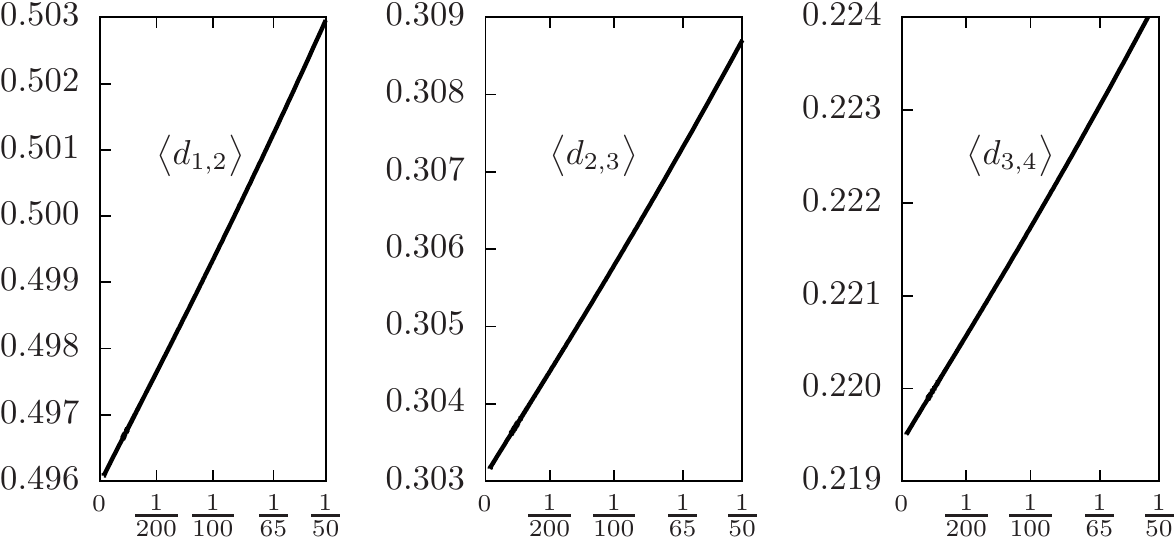}
\caption{The average distances between the first rightmost particles 
$\langle d_{1,2}(t)\rangle$, $\langle d_{2,3}(t)\rangle$ and $\langle
d_{3,4}(t)\rangle$ of a branching random walk versus $1/t$, for $t$ up to
3000.}\label{dn_of_t} 
\end{figure}

We have integrated numerically the equations satisfied by
$\psi_\lambda(x,t) $  and its derivatives with respect to $\lambda$ to
measure the distances $\langle d_{n,n+1}(t) \rangle $ between the
$n^\text{th}$ and $(n+1)^\text{th}$ rightmost particles. In our numerical
integration, we had to discretize space and time; we checked that our
results shown in figure \ref{dn_of_t} were stable when we  decreased our
integration steps.

One can first remark that, in contrast to standard Monte-Carlo simulations,
where all the branching events would be simulated and for which the
maximum reachable time would be $t\sim 20$ (with a number of particles
$e^t$ of order $10^{9}$), the integration of (\ref{Q0}) or of its
derivatives allows one to achieve much larger times. One can also notice
in figure \ref{dn_of_t} that the distances converge like $1/t$ to well
defined values. We will see that this $1/t$ convergence is consistent with
our analytic expression (\ref{Y-t}) below.

We did not find  an analytic theory to predict  the limiting values
that we  measured as in figure~\ref{dn_of_t}:
\begin{equation}\begin{aligned}
 \langle d_{1,2}\rangle&\simeq0.496,
&\langle d_{2,3}\rangle&\simeq0.303,
&\langle d_{3,4}\rangle&\simeq0.219,\\
 \langle d_{4,5}\rangle&\simeq0.172,
&\langle d_{5,6}\rangle&\simeq0.142,
&\langle d_{6,7}\rangle&\simeq0.121.
\label{valuesdn}
\end{aligned}\end{equation}
As shown  below (\ref{dn-large-n}),
 we can however predict their large $n$ behavior.

Before doing so, it is interesting to compare our results (\ref{valuesdn})
to the expected values of the gaps between the low lying energies of
spin-glass models such as the REM and the GREM
\cite{Derrida.81,Derrida.85}. In these models one can
show that these energies are given by probability cascades
\cite{Ruelle.87, Aizenman.05, Aizenman.07, Arguin.07, BovierKurkova.06}
and that the energy gaps at the leading edge are the same as
those of a Poisson process on the line with an exponential density. For
such a process, with density $e^{-\alpha x}$, the probability distribution
of the positions is $p_n(x)= \exp[- n \alpha x - e^{-\alpha x}/\alpha]$
from which one gets through (\ref{xn},\ref{dn})
\begin{equation}
\langle d_{n,n+1}\rangle_\text{GREM} = {1 \over \alpha   n}.
\label{Poisson-process}
\end{equation}
Clearly there is no choice of $\alpha$ for which our  numerical results
(\ref{valuesdn}) are  compatible with (\ref{Poisson-process}).

It is well known \cite{Bramson.78, vanSaarloos.03}  that the solution
$Q_0(x,t)$ of the Fisher-KPP equation (\ref{Q0}) with the step initial
condition (\ref{initial-condition}) becomes, for  large $t$, a traveling
wave of the form
\begin{equation}
Q_0(x,t) \simeq F[x-\langle X_1(t)\rangle],
\label{trav}
\end{equation}
where the shape $F(z)$ of the front  [$F(z) \to 1$ as $z \to \infty$ and
$F(z) \to 0$ as $z \to -\infty$] is time-independent and its position,
which can be defined as the average position  $\langle X_1(t) \rangle$ of
the rightmost particle, has the following long time behavior
\cite{Bramson.78, BrunetDerrida.97, vanSaarloos.03}
\begin{equation}
 \langle X_1(t) \rangle = 2 t - {3 \over 2} \ln t + {\cal O}(1).
\label{X1}
\end{equation}
$\psi_\lambda(x,t)$ is also the solution of the Fisher-KPP equation
(\ref{Q0}), but with the initial condition (\ref{initial-condition-bis}).
As $\psi_\lambda(x,0)$ decays fast enough \cite{vanSaarloos.03}, one
expects the same large $t$ behavior as (\ref{trav},\ref{X1}), up to
a $\lambda$-dependent delay $f(\lambda)$ due to the change of initial
condition:
\begin{equation}
\psi_\lambda(x,t) \simeq F[x-\langle X_1(t)\rangle + f(\lambda)].
\label{trav-1}
\end{equation}
From (\ref{retard-1},\ref{trav},\ref{trav-1})  we see that the translation
$f(\lambda)$ is nothing but the generating function of the average distances
\begin{equation}
f(\lambda) = \lim_{t \to \infty} 
\sum_{n \geq 1}  \lambda^n \langle d_{n,n+1}(t) \rangle.
\label{retard-2}
\end{equation}

We were not able to find an analytic expression of the delay function
$f(\lambda)$ for arbitrary $\lambda$. For $\lambda$ close  to $1$, however,
we are going to show that
\begin{equation}
f(\lambda) = \tau_\lambda    -\ln \tau_\lambda  +{\cal O}(1)  \quad \text{with }
\tau_\lambda= - \ln(1- \lambda).
\label{retard-3}
\end{equation}
This implies (\ref{retard-2}) that the distances  have the following large
$n$ asymptotics
\begin{equation}
\langle d_{n,n+1}(t) \rangle_{t=\infty} \simeq {1 \over n} - {1 \over n
 \ln n} + \cdots
\label{dn-large-n}
\end{equation}
Compared with  (\ref{Poisson-process}), we see that there is a correction,
which  we believe to be universal as discussed below. (Note that the same
asymptotic distances would be obtained for uncorrelated particles
distributed according to a Poisson point process
with a density $-x e^{-x}$ for negative~$x$.)

For $\lambda$ close to $1$, the time $\tau_\lambda$ in \eqref{retard-3} is
the characteristic time it takes $\psi_\lambda(-\infty,t)$ to reach a value
close to~0. The most na\"ive idea to derive \eqref{retard-3} would be to
say that it takes this time $\tau_\lambda$ for $\psi_\lambda(x,t)$ to
look like the
step function $Q_0(x,0)$, and then to start moving like $Q_0(x,t)$. As the
asymptotic velocity is $2$ (see (\ref{X1})) this would lead to a delay
$f(\lambda) \simeq  2 \tau_\lambda$ which is wrong (see (\ref{retard-3}))
by a factor $2$. The problem with this idea is that, while
$\psi_\lambda(x,t)$ evolves to approach $0$ on the negative $x$ axis,
a tail builds up on the positive $x$ axis which has a strong influence on
the dynamics later on.

To  derive \eqref{retard-3}, we need to understand the shape of
$\psi_\lambda(x,t)$ for $t>\tau_\lambda$. Let $Y_t$ be the position where
$\psi_\lambda(Y_t,t)=1/2$. We  have checked both numerically and
analytically that the following picture holds for $t$ and $\tau_\lambda$
large, with a given ratio $t/\tau_\lambda$ larger than 1:

In the range where $x-Y_t$ is  of order~$1$   
\begin{equation}
\psi_\lambda(x,t) \simeq \phi_{v(t)}(x-Y_t),
\label{defphi-v}
\end{equation}
where  $v(t)=\dot{Y_t}$ is the instaneous velocity of the front and
$\phi_v$ is the solution of the Fisher-KPP equation moving at  a constant
velocity $v$,
i.e. the solution of
\begin{equation}
\phi_v'' + v \phi_v' + \phi_v^2 - \phi_v=0,
\label{phi-v}
\end{equation}
with $\phi_v(-\infty)=0$ and $\phi_v(+\infty)=1$.
(The same form \eqref{defphi-v} is used in \cite{EbertvanSaarloos.98}.)
For definiteness, we normalize such that $\phi_v(0)=1/2$. This
determines a unique solution which has, if $v>2$, the
following asymptotics  for $z \to + \infty$:
\begin{equation}
1-\phi_v(z)  \simeq B_\gamma \ e^{-\gamma z}  + o( e^{-\gamma z} ),
\label{phi-v-asympt}
\end{equation}
where $\gamma$ is the  smallest solution of
\begin{equation}
v= \gamma + \gamma^{-1}.
\label{v(gamma)}
\end{equation}

On the other hand, in  the range $x-Y_t \gg 1$,
$\psi_\lambda(x,t)$ is accurately given by the solution
of the equation obtained by linearizing   (\ref{Q0}) around $1$:
\begin{equation}
1- \psi_\lambda(x,t) \simeq {(1- \lambda) e^t \over \sqrt{\pi}}
\int_{x/\sqrt{4 t}}^\infty e^{-u^2} du.
\label{error}
\end{equation}
Then, using the asymptotics of the error function $\int_X^\infty
\exp({-u^2})
du \simeq \exp({-X^2})/(2 X)$ and requiring  that 
(\ref{defphi-v},\ref{phi-v-asympt})   and  (\ref{error}) match
in the range $1 \ll x-Y_t \ll \sqrt{t}$, one gets
that $\gamma(t)$ and $Y_t $ should satisfy 
\begin{equation}
B_{\gamma(t)}   e^{-\gamma(t)(x-Y_t)} \simeq {(1- \lambda)  e^t  \sqrt{t
}\over Y_t \sqrt{\pi }}    e^{-{Y_t^2 \over 4 t}- {Y_t       
 (x - Y_t) \over 2 t}}.
\end{equation}

\begin{figure}[ht]
\includegraphics[width=\columnwidth]{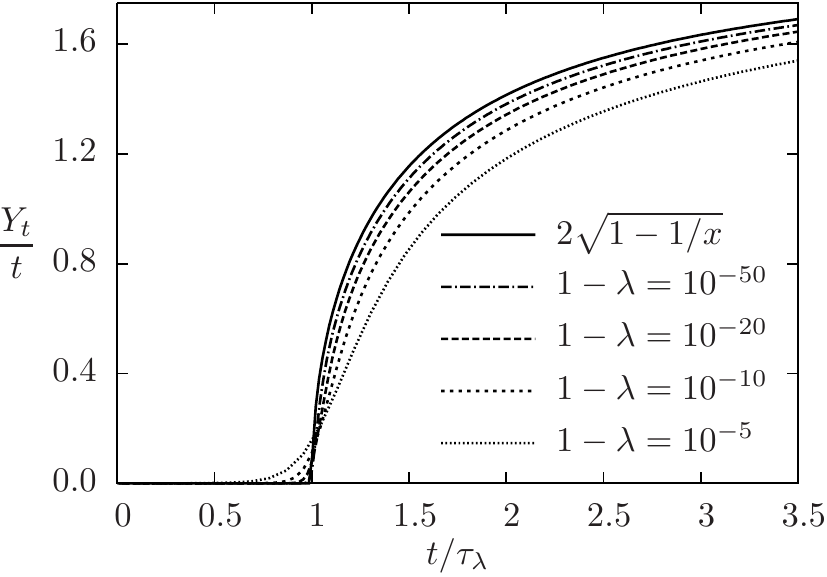}
\caption{The prediction (\ref{Y-t}) to the leading order for the position
of the front is compared to the position measured by integrating (\ref{Q0})
with the initial condition (\ref{initial-condition-bis}). As in
\eqref{retard-3}, $\tau_\lambda=-\ln(1-\lambda)$.}
\label{retard}
\end{figure}

To match  the dependence in $x-Y_t$, we need $Y_t\simeq 2 t \gamma(t)$ to
first order. Then matching the prefactors leads to
\begin{gather}
Y_t  \simeq 2 \sqrt{t (t-\tau_\lambda)} -  { \ln t  \over 2
\gamma(t) }-{\ln[ 2 \sqrt{\pi}  \gamma(t)  B_{\gamma(t)}  ] \over
\gamma(t)}, \label{Y-t}\\
\text{with}\qquad\gamma(t) \simeq \sqrt{t- \tau_\lambda \over t}.
\label{gamma-t}
\end{gather}
Note that the relation (\ref{v(gamma)}) is indeed satisfied to leading
order as $\dot{Y_t}\simeq\gamma(t)+\gamma(t)^{-1}$. In
figure \ref{retard} we see that  the agreement between
the leading term in (\ref{Y-t}) and the position obtained by integrating
numerically 
  (\ref{Q0}) with the initial condition
(\ref{initial-condition-bis}) is quite good.
 One could  also see in (\ref{Y-t}) a $1/t$  convergence of
the $\lambda$-dependent delay which is consistent with the numerical
results of figure  \ref{dn_of_t}.

For $t/\tau_\lambda$ large, $\gamma(t) \to 1 $ and $v(t) \to 2$. For $v=2$,
 the solution of (\ref{phi-v}) satisfies
$1-\phi_2(z) \simeq A z e^{-z}$   for large $z$
\cite{BrunetDerrida.97,vanSaarloos.03}. For $v$ slightly larger
than~2, the next term in the large~$z$ expansion \eqref{phi-v-asympt} is
$1-\phi_v(z)  \simeq B_\gamma \ e^{-\gamma z}  +  C_{\gamma}   e^{-z/\gamma }  +
o(e^{-z/\gamma })$. For consistency in the limit $v\to2$, one has
\begin{equation}
B_\gamma \simeq - C_\gamma \simeq {A \over 2 (1-\gamma)} \quad \text{as }
\gamma \to 1,
\label{consistent}
\end{equation}
so that $B_{\gamma(t)}\simeq At/\tau_\lambda$
from (\ref{gamma-t},\ref{consistent}). Thus, \eqref{Y-t} becomes
$Y_t\simeq2t-\tau_\lambda-(3/2)\ln t +\ln\tau_\lambda+{\cal
O}(1)$,  which gives
(\ref{X1},\ref{trav-1},\ref{retard-3}).

One can repeat everything if, instead of starting with a single particle at
the origin, one  starts with $K$ of particles at positions
$y_1,\ldots, y_K$.  One  simply needs to replace $\psi_\lambda(x,t)$
defined in (\ref{psi-def})  by $\prod_{1 \leq i \leq K}
\psi_\lambda(x-y_i,t)$, with a similar change for  $Q_0(x,t) \equiv
\psi_0(x,t)$. As a result, in the long time limit, the  delay function
$f(\lambda)$, and therefore the distances between the rightmost particles
remain unchanged. This property is remarkable: whatever the positions of
the initial particles are (as long as there are a finite number of them)
the limiting average distances and probably  the whole limiting measure
seen from the rightmost particle are the same.

We can also extend  all our calculations to more general branching random
walks. For example one may consider a discrete time case where at each time
step, every particle splits into $K$ new particles, and  the position of
each new particle is shifted from its parent by a random amount  $\epsilon$
drawn from a given  distribution $\rho(\epsilon)$. Apart from a few
changes, such as (\ref{v(gamma)}) which is replaced  by
\begin{equation}
\label{v-bis}
v = \gamma^{-1}\ln\left[ K  \int \rho(\epsilon)  \ e^{\gamma
\epsilon} \   d \epsilon \right],  
\end{equation}
$\tau_\lambda$ in (\ref{retard-3})
which becomes $-\ln(1-\lambda)/\ln K$  or
 $\gamma(t)$ in  (\ref{gamma-t}) 
which becomes the solution of 
\begin{equation}
\gamma^2 {d v \over d \gamma} = -{\ln(1-\lambda) \over t},
\end{equation}
everything remains unchanged. In particular 
(\ref{retard-3},\ref{dn-large-n}) are simply divided by the value
$\gamma_0$ of $\gamma$ which minimizes the expression (\ref{v-bis}) of
$v$.
Thus, the asymptotics of both the delay (\ref{retard-3}) and the distances
(\ref{dn-large-n}) look universal, up to a scale factor $\gamma_0$.

In the present letter we have seen that the distances between the
rightmost particles at the frontier of a branching random walk have
statistical properties (\ref{valuesdn},\ref{dn-large-n}) which can be
understood as the delay (\ref{trav-1},\ref{retard-2},\ref{retard-3}) of
a traveling wave. Other properties, such as the correlations of these
distances or even their whole probability
distribution can also be understood in terms of the delay of a traveling
wave. For example if $R_{n,m}(x,y,t)$ is the probability that there are $n$
particles at the right of $x$ and $m$ particles at the right of $y$,
on can show that
\begin{equation*}
\langle d_{n,n+1}(t) d_{m,m+1}(t) \rangle = \int dx\, x \int dy\,y{ \partial^2
R_{n,m}(x,y,t) \over \partial x \partial y},
\end{equation*}
while the generating function $\sum_{n,m}\lambda^n\mu^m R_{n,m}(x,x+c,t)$
defined as in (\ref{psi-def}) evolves according to the Fisher-KPP equation
(\ref{Q0}) with a new initial condition.

A surprising aspect of the present work is that the statistics of the
leading particles, in the long time limit, do not depend on the positions
or on the number of particles we start with, as long as there is a finite
number of them. This means that the limiting measure has the following
stability property: if one takes two realizations
of the leading particles according  to this measure and shifts one of
them by an arbitrary amount, then the superimposition of these two
realizations gives a new realization of the same measure, up to
a translation.


\end{document}